\documentclass[11pt]{article}
\usepackage{graphicx}
\usepackage{dcolumn}
\usepackage{amsmath}
\usepackage{pst-node}

\newcommand{\BABARPubYear}    {02}

\newcommand{\BABARConfNumber} {032}
\newcommand{\SLACPubNumber} {9310}

\input babarsym.tex
\input extradefs.tex

\long\def\inst#1{\par\nobreak\kern 4pt\nobreak
    {\it #1}\par\vskip 10pt plus 3pt minus 3pt}

\begin{document}
{\pagestyle{empty}

\begin{flushright}
%\babar\ Analysis Document \#437, Version 7 \\
\babar-CONF-\BABARPubYear/\BABARConfNumber \\
%\babar-PUB-\BABARPubYear/\BABARPubNumber \\
SLAC-PUB-\SLACPubNumber \\
%hep-ex/\LANLNumber \\
July 2002 \\
\end{flushright}

\par\vskip 5cm

% Title of the paper
\begin{center}
\Large \bf A Search for the Decay \boldmath{\Bztopizpiz} 
\end{center}
\bigskip

\begin{center}
\large The \babar\ Collaboration\\
\mbox{ }\\
July 24, 2002
\end{center}
\bigskip \bigskip

% Abstract
\begin{center}
\large \bf Abstract
\end{center}
We present a search for the decay \Bztopizpiz by the \babar\ 
experiment at the \pep2 asymmetric-energy B-factory at SLAC.  Using
approximately 88 million \BB pairs collected between 1999 and 2002, we
place a 90\% confidence level upper limit on the branching fraction of
\begin{displaymath}
\mathcal{B}(\Bztopizpiz) < 3.6 \times 10^{-6} \, .
\end{displaymath}
This result is preliminary.
\vfill
\begin{center}
Contributed to the 31$^{st}$ International Conference on High Energy Physics,\\ 
7/24---7/31/2002, Amsterdam, The Netherlands
\end{center}

\vspace{1.0cm}
\begin{center}
{\em Stanford Linear Accelerator Center, Stanford University, 
Stanford, CA 94309} \\ \vspace{0.1cm}\hrule\vspace{0.1cm}
Work supported in part by Department of Energy contract DE-AC03-76SF00515.
\end{center}

\newpage
} % end of pagestyle{empty}

% Input author list file
\begin{center}
\small

The \babar\ Collaboration,
\bigskip

%% author list as of 05-Jul-2002 (556 authors)
B.~Aubert,
D.~Boutigny,
J.-M.~Gaillard,
A.~Hicheur,
Y.~Karyotakis,
J.~P.~Lees,
P.~Robbe,
V.~Tisserand,
A.~Zghiche
\inst{Laboratoire de Physique des Particules, F-74941 Annecy-le-Vieux, France }
A.~Palano,
A.~Pompili
\inst{Universit\`a di Bari, Dipartimento di Fisica and INFN, I-70126 Bari, Italy }
% ...

\end{center}\newpage

%%%% Body %%%%%%%%%%%%%

% Theoretical introduction 
\section{Introduction}
\label{sec:Introduction}
The study of \B meson decays into charmless hadronic final states
plays an important role in the understanding of \CP violation in the
\B system.  Measurements of the \CP-violating asymmetry in the
\Bztopipi decay mode can provide information on the angle $\alpha$ of
the Unitarity Triangle. However, in contrast to the theoretically
clean determination of the angle $\beta$ in \B decays to charmonium
final states~\cite{babarsin2beta,bellesin2beta}, the extraction
of $\alpha$ in \Bztopipi is complicated by the interference of tree and
penguin amplitudes with different weak phases. The time dependent
\CP-violating asymmetry in \Bztopipi is proportional to
$\sin{2\alpha_{\rm eff}}$. 
Assuming an isospin
relation~\cite{GronauLondon}, 
$|\alpha_{\rm eff} - \alpha|$ may be determined from the branching
fractions $\mathcal{B}(\Btopipiz)$, $\mathcal{B}(\Bztopipi)$,
$\mathcal{B}(\Bzbtopipi)$,   $\mathcal{B}(\Bztopizpiz)$, and
$\mathcal{B}(\Bzbartopizpiz)$. Alternatively, a bound on $\alpha_{\rm
  eff} - \alpha$ may be found from the ratio
$\mathcal{B}(\Bztopizpiz)/\mathcal{B}(\Btopipiz)$, using the average
of \Bz and \Bzb branching fractions~\cite{GQ}.
In this paper, we report on a search for the decay \Bztopizpiz.  Here
and throughout this paper \Bztopizpiz is meant to include both \Bz and
\Bzb decays.  

% Data sample 
\section{The \babar\ Detector and Dataset}
\label{sec:babar}

\babar\ is a solenoidal detector optimized for the asymmetric beams at
\pep2 and is described in detail
in Ref.~\cite{babarnim}. Charged particle (track) momentum and
direction are measured with a 5-layer double-sided silicon vertex
tracker (SVT) and a 40-layer  drift chamber (DCH)
embedded in a 1.5 T superconducting solenoidal magnet. 
Neutral cluster position and energy are measured by an electromagnetic
calorimeter (EMC) consisting of 6580 CsI(Tl) crystals. The photon
energy resolution is $\sigma_{E}/E = (2.32 / E(\gev)^{1/4} \oplus 1.85)
\%$, and the angular resolution is $\sigma_{\theta} =
3.87^{\rm o}/\sqrt{E(\gev)}$.  Charged hadrons are identified  with a
detector of internally reflected Cherenkov light (DIRC) and specific
ionization in the tracking detectors. The
instrumented magnetic  flux return (IFR) detects neutral hadrons
and identifies muons.

This search uses $(87.9\pm 1.0) \times 10^{6}$ \BB pairs from
approximately $81\invfb$ of data at the \FourS resonance
(on-resonance), and approximately $9\invfb$ of data at $40\mev$ below
the \FourS resonance (off-resonance), collected with the \babar\ 
detector from 1999 through 2002.  The \pep2 collider is operated with
asymmetric beam energies, corresponding to a boost for the \FourS of
$\beta\gamma = 0.55$.

\section{Event Selection}
\label{sec:Event}

% Event selection cuts 
\BB events are selected using track and neutral cluster content and
event topology. Events are required to have either three or more well
measured tracks from the interaction point with transverse momentum
$\pt > 0.1~\gevc$ and polar angle in the lab frame $0.41 < \theta_{\rm
  lab} < 2.54 $~rad, or two or fewer such tracks combined with two or
more neutral clusters with center-of-mass (CM) energy $E_{CM} >
0.5\gev$ and one or more additional neutral clusters with laboratory
energy $E_{\rm lab} > 0.1\gev$.  Backgrounds from lepton pair events
are removed by requiring that the ratio of the second to zeroth
Fox-Wolfram moment be less than $0.95$ and the event sphericity be
greater than $0.01$.  The principal background is from the $\epem \to
\qqbar$ process ($q=u, d, s, c$), when both quark jets contain a \piz
which combine to mimic a \B decay. This background is suppressed by
requiring that the cosine of the angle between the sphericity axis of
the \B candidate and the sphericity axis of the remaining tracks and
neutral clusters in the event satisfy $|\cos{\theta_{S}}| < 0.7$.

\section{Candidate Selection}
\label{sec:Candidate}

Candidate \piz mesons are formed from two neutral clusters with $E >
0.03 \gev$ whose transverse energy profile in the EMC is consistent
with that of a single photon. The centroid of the two clusters must be
separated by at least one EMC crystal.  To reduce the background from
false \piz candidates, the cosine of the angle between the $\gamma$
momentum vector in the \piz rest frame and the \piz momentum vector in
the lab frame is required to satisfy $|\cos{\theta_{\gamma}}| < 0.95$.
The invariant mass of the two photons is required to be within $\pm 3
\sigma$ of the \piz mass.

% pi0pi0
 
\Bztopizpiz candidates are formed from pairs of \piz candidates.  The remaining
background is from \qqbar events that have a spherical topology and
pass the $|\cos{\theta_{S}}|$ requirement, and \Btorhopiz decays in
which the  $\pipm$ is emitted nearly at rest in the \B frame.
No other \B decay produces a significant background for \Bztopizpiz. The
\Btorhopiz decay mode has not been observed; the limit on its
branching fraction is $\mathcal{B}(\Btorhopiz) < 4.3 \times 10^{-5}$ at 90\%
CL~\cite{cleorhopi}.  Both backgrounds are separated from signal by
using the kinematic constraints of \B mesons produced at the \FourS. The
first kinematic parameter is a beam-energy substituted mass $\mes =
\sqrt{E^{2}_{b} - {\bf p}^{2}_{B}}$, where $E_{b} = (s/2 + {\bf
  p}_{i}\cdot{\bf p}_{B})/E_{i}$; $\sqrt{s}$ and $E_{i}$ are the total
energy of the \epem system in the CM and laboratory frames,
respectively, and ${\bf p}_{i}$ and ${\bf p}_{B}$ are the momentum
vectors in the lab frame of the \epem system and the \B candidate,
respectively.  The second kinematic parameter is $\de = E_{B} -
\sqrt{s}/2$, where $E_{B}$ is the \B candidate energy calculated in
the CM frame.  In \Bztopizpiz events the \mes and \deltaE resolution
are predicted to be approximately $ 3.8\mevcc$ and $ 80\mev$,
respectively, based on simulation.

% pi0pi0

There are 3020 candidates with $\mes > 5.2\gevcc$ and $|\de| < 0.2\gev$
which are used in this search.  The \Bztopizpiz signal efficiency is
evaluated with a GEANT4 based detector simulation~\cite{GEANT4}.  The
efficiency to separate closely spaced photons in the EMC is measured using
$\tau^{\pm}\to\pi^{\pm}\piz\nu_{\tau}$ and
$\tau^{\pm}\to\pi^{\pm}\piz\piz\nu_{\tau}$ decays, and uncertainty in
this efficiency dominates the error in the signal efficiency.  The
\Bztopizpiz efficiency is $(16.5 \pm 1.7)\%$.
  
The \Btorhopiz background is reduced by removing candidates in
which  the omitted \pipm is identified.  Tracks that are not
identified as leptons or kaons, and that are not part of a
reconstructed $\KS\to\pip\pim$, $\Lambda \to p \pi$, or
$\gamma\to\epem$ candidate,
are used.  The track that has a $\pipm\piz$ invariant mass and
$\mes$ of the $\pipm\piz\piz$ combination most consistent with the
$\rho$ mass and \Btorhopiz hypothesis is selected.  A
cut is applied on a linear combination of the $\pipm\piz$ invariant
mass  and the $\deltaE$ of the $\pipm\piz\piz$ combination which removes roughly 50\% of
\Btorhopiz, with 93\% efficiency for \Bztopizpiz. 
Only $(0.40 \pm 0.04)\%$ of \Btorhopiz decays remain after all cuts.

The \qqbar background that remains after all cuts is further
distinguished from signal using a Fisher discriminant
$\mathcal{F}_{T}$ that combines energy flow and \B flavor tagging
variables.  The energy flow variables are $L_0 = \sum_{i} p_i$, and
$L_2 = \sum_{i} p_i \times \frac{1}{2}(3 \cos^2{(\theta_i)}-1)$, where
the sum is over all tracks and neutral clusters in the event except
the daughters of the \Bztopizpiz candidate. Here $\theta_{i}$ is the
angle with respect to the thrust axis of the \B candidate and $p_{i}$
is the momentum magnitude, both in the CM frame.  The \B flavor
tagging variable is a quality index which classifies the lepton,
charged kaon, and slow pion $\pipm_{\rm slow}$ (from the decay
$\Dstarpm\to\Dz\pipm_{\rm slow}$) content of the event. The quality
index is ordered by the degree of background rejection.  The leptons,
charged kaons, and slow pions are selected and the events are
classified with the \B flavor tagging algorithm described in
Ref.~\cite{babarsin2beta}. The coefficients of $\mathcal{F}_{T}$ are
optimized using Monte Carlo simulation of signal and \qqbar
background.

\begin{figure}[!tbh]
\begin{center}
\includegraphics[width=0.85\linewidth]{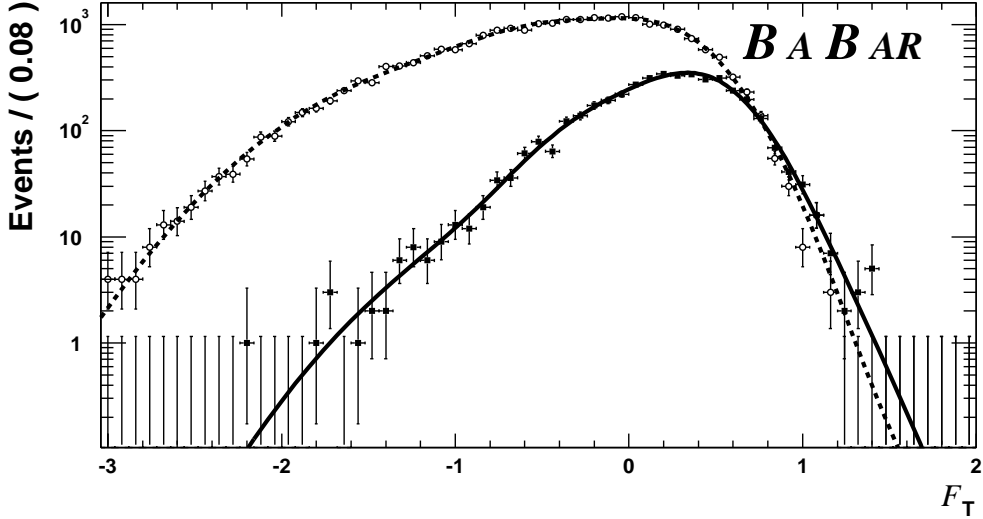}
\caption{
  The distribution of the Fisher discriminant $\mathcal{F}_{T}$ from a
  fully reconstructed $\B \to D^{(*)}n\pi$ data sample (open circles), and from
  off-resonance data and on-resonance \mes sidebands (filled squares).
  The triple Gaussian parameterizations used in the likelihood fit for
  \Bztopizpiz signal (dotted line) and \qqbar background (solid line)
  are also shown.  }
\label{fig:fisher}
\end{center}
\end{figure}

%Likelihood
\section{Unbinned Maximum Likelihood Fit}
\label{sec:fits}

The number of \Bztopizpiz events is determined by an unbinned extended maximum
likelihood fit to \mes, \de, and $\mathcal{F}_{T}$.
The probability ${\cal P}_i\left(\vec{x}_j; \vec{\alpha}_i\right)$ for a given hypothesis
is the product of probability density functions (PDFs) for each of the
variables $\vec{x}_j=(\mes,\de,\mathcal{F}_{T})$ given the set of parameters $\vec{\alpha}_i$.
The likelihood function is given by a product over all events $N$ and
three signal and background components:
\begin{displaymath}
{\cal L}= \exp\left(-\sum_{i=1}^3 n_i\right)\,
\prod_{j=1}^N \left[\sum_{i=1}^3 n_i {\cal P}_i\left(\vec{x}_j;
\vec{\alpha}_i\right)
\right]\, .
\end{displaymath}
The $n_i$ are the number of events in each of the three components: 
\Bztopizpiz  ($n_{\piz\piz}$), \Btorhopiz
($n_{\rho\piz}$),
and \qqbar ($n_{\qqbar}$).
Monte Carlo simulations are used to verify that the fit is unbiased.

% Pdfs

The \mes PDF for \qqbar is parameterized by a threshold function~\cite{argus}
\begin{displaymath}
f(\mes) = \mes\sqrt{1-(\mes/m_{0})^2}\exp{\left\{-\xi(1-(\mes/m_{0})^2)\right\}},
\end{displaymath}
where $m_{0}$ is the average CM beam energy, and  $\xi$ is 
found from a fit to on-resonance data with $|\cos{\theta_{S}}| > 0.9$.
The \deltaE PDF for \qqbar is parameterized by a quadratic function with
coefficients found from a fit to both on-resonance
data in the \mes sidebands and off-resonance data.  The 
$\mathcal{F}_{T}$ PDF for \qqbar is the sum of three Gaussians and is
also found using both \mes sideband and off-resonance data, as shown in
Fig.~\ref{fig:fisher}.  The \mes and \de PDFs for signal and
\Btorhopiz background are found from Monte Carlo simulation. The
\Bztopizpiz and \Btorhopiz \mes and \de variables are correlated, so a
two dimensional PDF derived from a smoothed Monte Carlo distribution
is used.  The $\mathcal{F}_{T}$ PDFs, shown in Fig.~\ref{fig:fisher}, for both \Bztopizpiz and
\Btorhopiz are parameterized as the sum of three Gaussians and are found
from a sample of fully reconstructed $\Bz \to D^{(*)} n\pi$ events,
with $n =1$, $2$, or $3$.

%\section{Results}
%\label{sec:Physics}

% Results and Systematics for h+pi0 

\begin{figure}[!tbh]
\begin{center}
\includegraphics[width=0.49\linewidth]{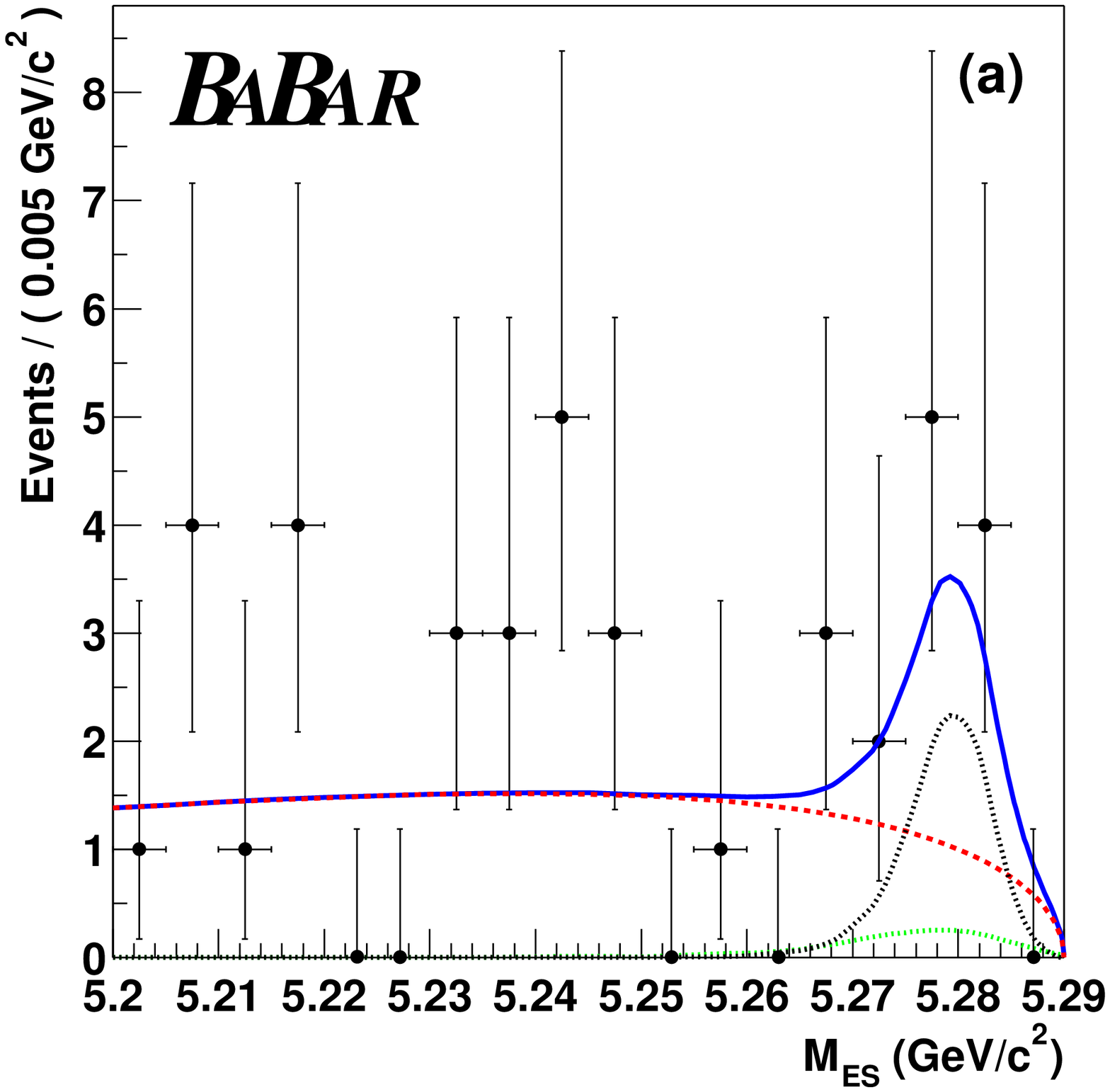}
\includegraphics[width=0.49\linewidth]{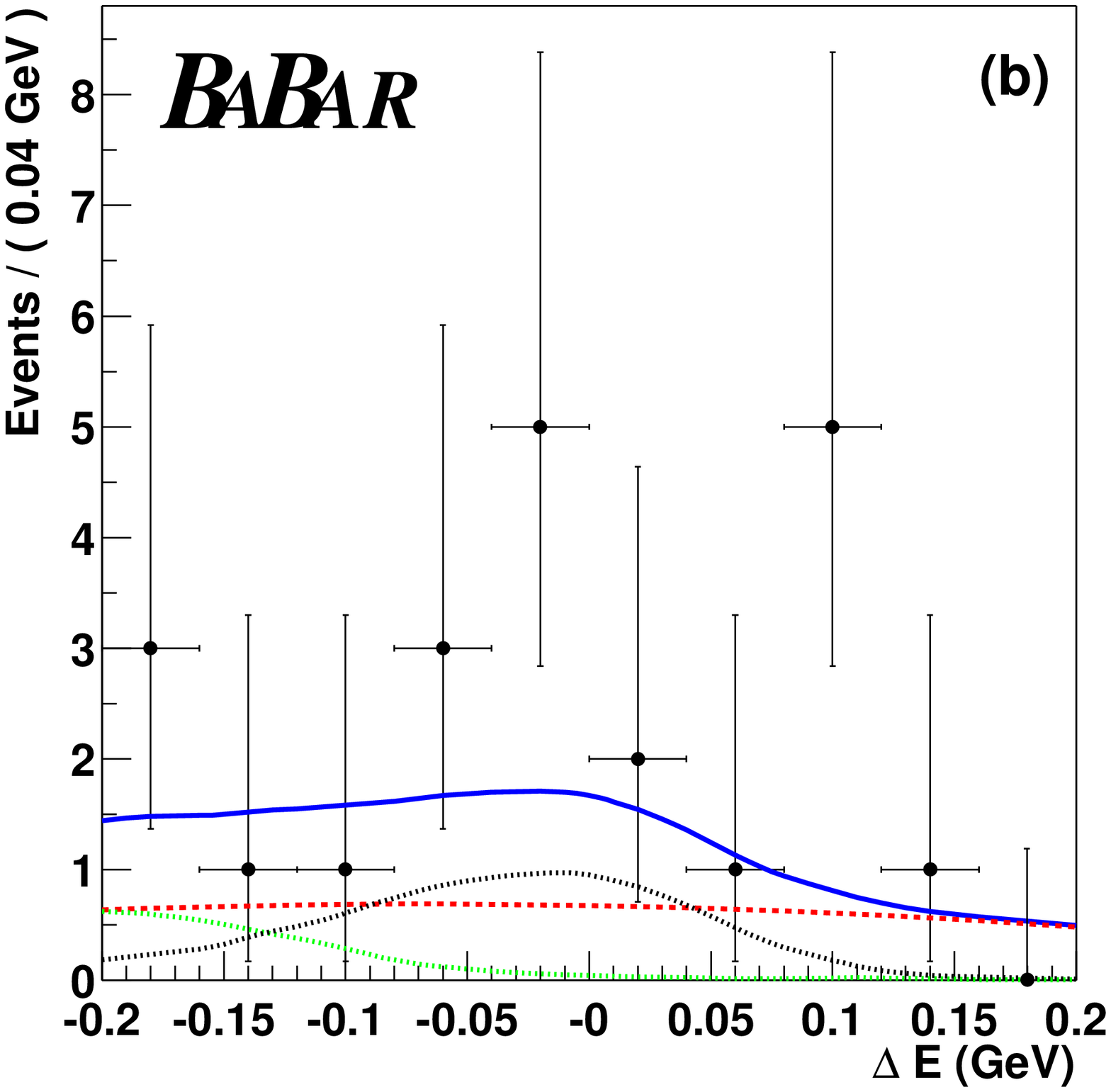}
\includegraphics[width=0.49\linewidth]{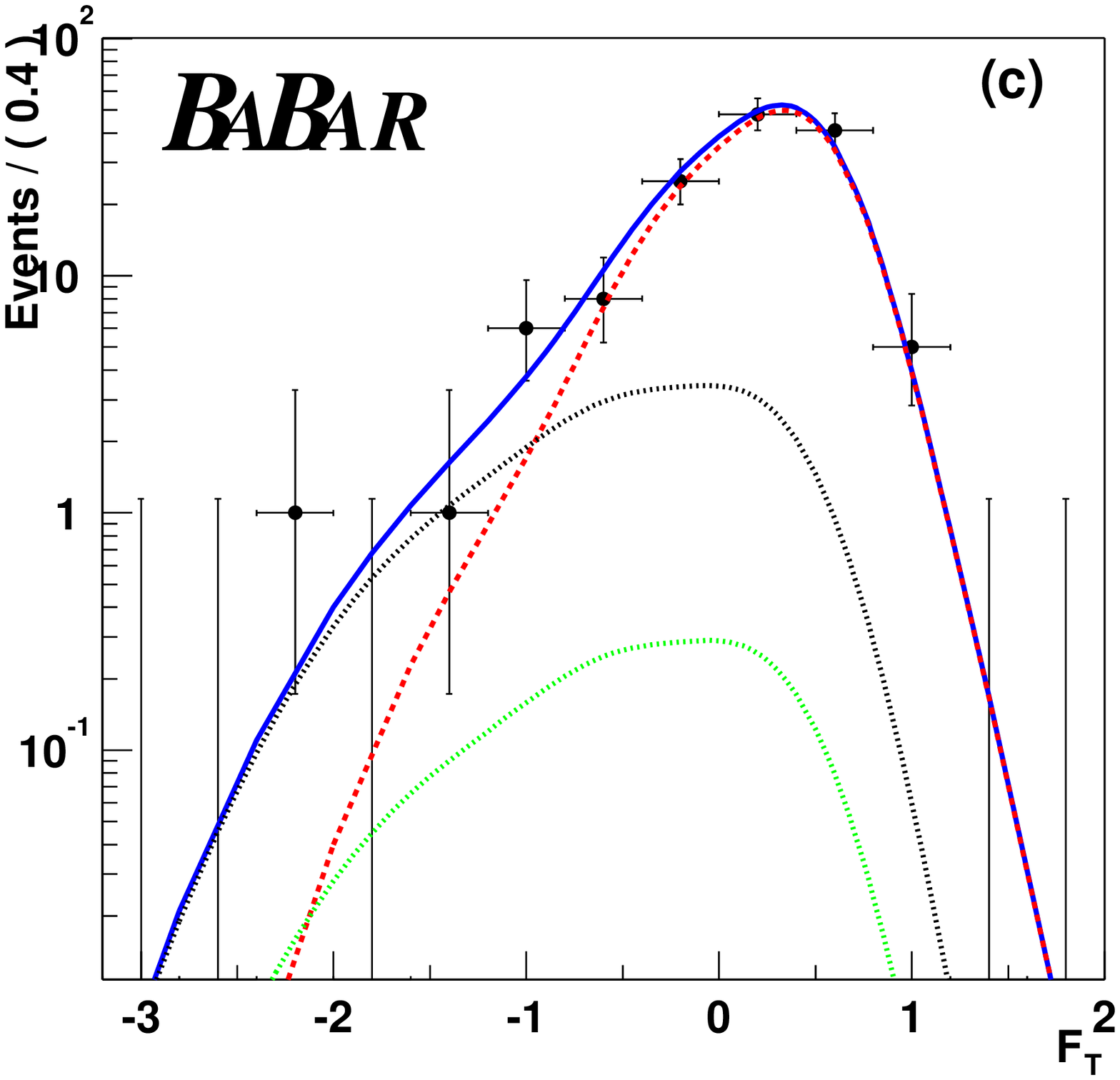}
\includegraphics[width=0.49\linewidth]{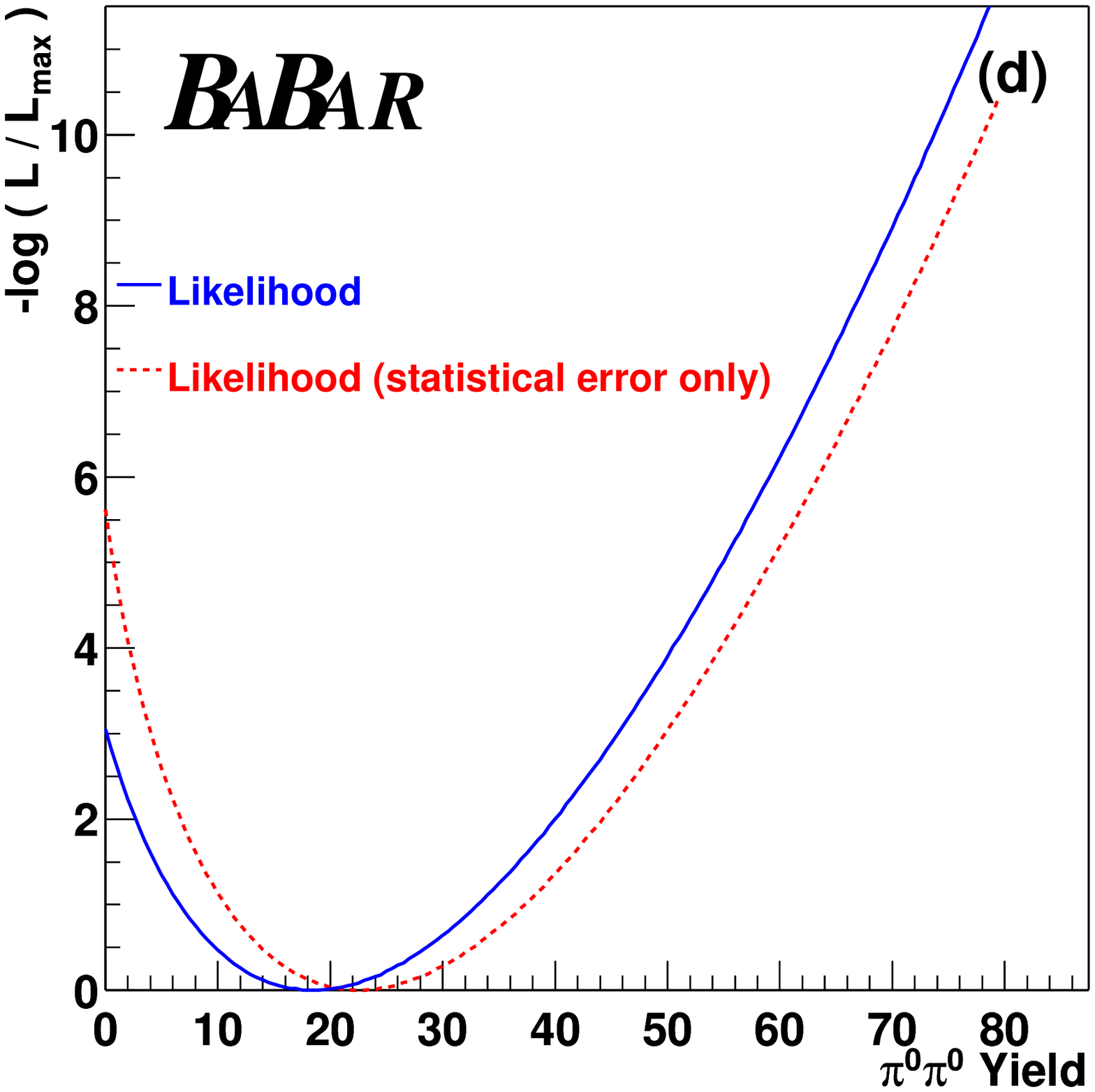}
\caption{Results from the maximum likelihood fit.   The distributions
for a) \mes, b) \de, c) $\mathcal{F}_{T}$ are shown, for  candidates
 satisfying an optimized cut on the probability ratio  $\mathcal{R}$. Also shown are the
PDF projections for signal (dotted line), \Btorhopiz (dot-dashed
line), \qqbar background (dashed line), and the sum (solid
line). These plots do not represent the full information used in the
maximum likelihood fit, but only a subset of the data. The ratio
$-\log{(\mathcal{L}/\mathcal{L}_{max})}$ is shown in d) (solid line) and
with statistical errors only (dashed line).
}
\label{fig:fit}
\end{center}
\end{figure}

The result of the fit is $n_{\piz\piz} = 23^{+10}_{-9}$ and
$n_{\qqbar} = 2990\pm 55$ events.  These statistical errors correspond to
the point at which $\log{\mathcal{L}}$ changes by one half.
The number of \Btorhopiz events is fixed in the fit to 
$n_{\rho\piz} = 8.4$, based on the central value from
Ref.~\cite{cleorhopi} of
$\mathcal{B}(\Btorhopiz) = 2.4\times10^{-5}$ and our estimated efficiency.  
The distributions of \mes, \de, and $\mathcal{F}_{T}$ are shown in
Fig.~\ref{fig:fit} after a cut on the probability ratio
\begin{displaymath}
\mathcal{R} = \frac{n_{\piz\piz}\mathcal{P}_{\piz\piz}}
{n_{\piz\piz}\mathcal{P}_{\piz\piz} +
  n_{\rho\piz}\mathcal{P}_{\rho\piz} + n_{
    \qqbar}\mathcal{P}_{\qqbar}}.
\end{displaymath}
Here the $\mathcal{P}_{i}$ are products of the PDFs for the two other
variables, and the $n_{i}$ are the central values from the fit. The cut
is optimized by maximizing the ratio
\begin{displaymath}
  \mathcal{S} = \frac{n_{\piz\piz}\epsilon_{\piz\piz}}{\sqrt{n_{\piz\piz}\epsilon_{\piz\piz} + n_{\rho\piz}\epsilon_{\rho\piz} + n_{\qqbar}\epsilon_{\qqbar}}},
\end{displaymath}
where $\epsilon_{i}$ is the efficiency of the cut. The efficiencies
for the \mes distribution are 20\%, 12\%, and 0.8\% for the
\Bztopizpiz, \Btorhopiz, and \qqbar components, respectively.
The PDF projections for each of the fit components, scaled by the appropriate $\epsilon_{i}$, are
also shown in Fig.~\ref{fig:fit}.

The results from the likelihood fit are compared to an
analysis that simply uses  the number of candidates
satisfying the requirements  $ 5.260 < \mes < 5.289 \gevcc$, $ -0.16
< \de < 0.10 \gev$, and $\mathcal{F}_{T} < -0.20$.  These cuts were
chosen in advance by maximizing the ratio
\begin{displaymath}
\frac{N_{\piz\piz}}{\sqrt{N_{\piz\piz}+N_{\rho\piz}+N_{\qqbar}}} ,
\end{displaymath}
where $N$ is the number of events from each source that satisfy the
cuts.  There are 89 events satisfying these requirements. The number
of background \qqbar events was determined by scaling the number of
events with $ 5.20 < \mes < 5.26 \gevcc$, $ -0.16 < \de < 0.10 \gev$,
and $\mathcal{F}_{T} < -0.20$ by the appropriate factor given the
threshold function describing the \mes distribution.  The number of
background \Btorhopiz events was estimated using the efficiency from
the simulation. We find $N_{\piz\piz} = -6 \pm 11~(\text{stat.})$.
Using simulations based on our PDFs, and assuming a flat prior
distribution for $\mathcal{B}(\Bztopizpiz)$, we estimate that there is
a 2.5\% probability to observe 89 or fewer events given the central
value of our likelihood fit.

\section{Systematic Uncertainties}

We have estimated the systematic uncertainty in the likelihood fit by
varying the PDF parameters by their statistical errors, by using
different parametrizations, and by varying the \Btorhopiz branching
fraction.  In each case the likelihood fit to the data is repeated and
the change in $n_{\piz\piz}$ is used as the systematic uncertainty.
The systematic errors are listed in Table~\ref{tbl:syst}.
The dominant systematic uncertainty is due to the statistically limited
sample of data used to parameterize the $\mathcal{F}_{T}$ PDF for
\qqbar.  Since the parameters in the triple Gaussian are highly
correlated we transform to the uncorrelated parameter space and vary
the uncorrelated parameters by $\pm1\sigma$.  The fit is repeated for
each $1\sigma$ variation of the uncorrelated parameters, and the
positive and negative changes in $n_{\piz\piz}$ are separately summed
in quadrature.  The fit is also repeated using an interpolated
histogram as the \qqbar $\mathcal{F}_{T}$ PDF, with a change of
$\Delta(n_{\piz\piz}) = -1.1$ events.  The fit is repeated for values
for the \qqbar \mes shape parameter of $\xi = 24.3 \pm 1.3$, based on
the change in $\xi$ as a function of $\cos{\theta_{S}}$. The \qqbar
\de parameters are varied by their statistical errors.  The EMC energy
scale is varied by $\pm 10.4\mev$ based on the statistical error in
the mean of \de in the \Btohpiz analysis~\cite{babarpipiz}, and the
\Bztopizpiz \de PDF is changed accordingly.
The \Btorhopiz veto cut is varied and the
changes taken as a systematic error. Lastly, the \Btorhopiz branching
fraction is varied from $1.2\times10^{-5}$ to $4.3\times10^{-5}$.

\begin{table}[!htb]
\caption{Systematic errors on the number of \Bztopizpiz events in the
  maximum likelihood fit.  $\Delta_{\pm}(n_{\piz\piz})$ are the
  positive and negative change
  in the number of signal events from the likelihood fit for
  each systematic source.}
\begin{center}
\begin{tabular}{lcc} \hline\hline
Systematic                                    & $\Delta_{+}(n_{\piz\piz})$ (events)   
                                              & $\Delta_{-}(n_{\piz\piz})$ (events)     \\ \hline
\qqbar  $\mathcal{F}_{T}$ PDF parameters      & +7.5                         & $-2.4$     \\
\qqbar  $\mathcal{F}_{T}$ PDF functional form & +1.1                         & $-1.1$     \\
\qqbar \mes PDF                               & +1.2                         & $-1.1$     \\
\qqbar \de PDF                                & +1.0                         & $-0.2$     \\
\Bztopizpiz \de                               & +0.8                         & $-1.1$     \\
\Btorhopiz cut variation                      & +1.3                         & $-1.3$     \\ 
\Btorhopiz branching fraction                 & +1.6                         & $-1.9$     \\ \hline
Total systematic error on  $n_{\piz\piz}$     & +8.1                         & $-3.8$     \\ \hline 
Efficiency systematics                        & $10.1\%$                       & $-10.1\%$  \\ \hline
Total systematic                              & +8.4                         & $-4.4$     \\ \hline \hline
\end{tabular}
\end{center}
\label{tbl:syst}
\end{table}

We calculate the significance of the result, defined as $s =
\sqrt{-2\log{(\mathcal{L}(n_{\piz\piz}=0)/\mathcal{L}_{max})}}$, and
the 90\% CL upper limit. The upper limit is evaluated by finding $n^{UL}_{\piz\piz}$ where
\begin{displaymath}
\frac{\int_{0}^{n^{UL}_{\piz\piz}}\mathcal{L}(n) dn}{\int_{0}^{\infty}\mathcal{L}(n) dn} = 0.9 .
\end{displaymath}
 Systematic errors are included
in the following way.  For the significance, we repeat the fit using
the changes in \qqbar $\mathcal{F}_{T}$ parameterization, \qqbar \mes
parameterization, and \Btorhopiz branching fraction which cause
$n_{\piz\piz}$ to decrease.  The
$-\log{(\mathcal{L}/\mathcal{L}_{max})}$ function is shown in
Fig.~\ref{fig:fit}d, along with the same function before systematic
errors are included.  The significance of the result is $s=2.5\sigma$.
The systematic errors are included by adding the total systematic
$\Delta_{+}(n_{\piz\piz})$, in Table~\ref{tbl:syst},  to $n^{UL}_{\piz\piz}$.  We find
$n_{\piz\piz} < 46$~events at 90\% CL.

% Conclusions 
\section{Results}
\label{sec:conclusions}

To convert the number of events $n_{\piz\piz}$ into a branching
fraction we use
\begin{displaymath}
 \mathcal{B}(\Bztopizpiz) = 
 \frac{n_{\piz\piz}}{\epsilon_{\piz\piz} \cdot N_{\BB}}.
\end{displaymath}
$N_{\BB}=(87.9\pm1.0)\times10^{6}$ is the number of \BB pairs in our
data sample and the
efficiency is $\epsilon_{\piz\piz} = 0.165 \pm 0.017$.  The central
value of the likelihood fit is $\mathcal{B}(\Bztopizpiz) =
(1.6^{+0.7}_{-0.6} ({\rm stat.}) \, ^{+0.6}_{-0.3}
\textrm({\rm syst.}))\times10^{-6}$.  To calculate the branching fraction
upper limit we decrease  $\epsilon_{\piz\piz}$  and   $N_{\BB}$ 
by one $\sigma$. The upper limit on the branching fraction is 
\begin{displaymath}
  \mathcal{B}(\Bztopizpiz) < 3.6 \times 10^{-6} \text{\; at 90\% CL}.
\end{displaymath}
These results are preliminary.
The upper limit may be combined with our measurement of the branching
fraction $\mathcal{B}(\Btopipiz) = (5.5 \pm 1.0 \pm 0.6)\times
10^{-6}$~\cite{babarpipiz} to bound the ratio $
\mathcal{B}(\Bztopizpiz)/\mathcal{B}(\Btopipiz)$. Treating the
systematic uncertainties in the same way as for the
$\mathcal{B}(\Bztopizpiz)$
upper limit, and removing correlated
systematic uncertainties, we find $
\mathcal{B}(\Bztopizpiz)/\mathcal{B}(\Btopipiz) < 0.61$ at 90\%
CL. Assuming the isospin relations for $\Btopipi$~\cite{GQ} this corresponds to
an upper limit of $|\alpha_{\rm eff} - \alpha| < 51^{\rm o}$ at 90\% CL.

\section{Acknowledgments}
\label{sec:Acknowledgments}
% Standard acknowledgments paragraph; must always be included.
We are grateful for the 
extraordinary contributions of our \pep2\ colleagues in
achieving the excellent luminosity and machine conditions
that have made this work possible.
The success of this project also relies critically on the 
expertise and dedication of the computing organizations that 
support \babar.
The collaborating institutions wish to thank 
SLAC for its support and the kind hospitality extended to them. 
This work is supported by the
US Department of Energy
and National Science Foundation, the
Natural Sciences and Engineering Research Council (Canada),
Institute of High Energy Physics (China), the
Commissariat \`a l'Energie Atomique and
Institut National de Physique Nucl\'eaire et de Physique des Particules
(France), the
Bundesministerium f\"ur Bildung und Forschung and
Deutsche Forschungsgemeinschaft
(Germany), the
Istituto Nazionale di Fisica Nucleare (Italy),
the Research Council of Norway, the
Ministry of Science and Technology of the Russian Federation, and the
Particle Physics and Astronomy Research Council (United Kingdom). 
Individuals have received support from 
the A. P. Sloan Foundation, 
the Research Corporation,
and the Alexander von Humboldt Foundation.

\end{document}